\documentclass[aps,prb,twocolumn,groupedaddress,showpacs,floatfix]{revtex4}
\usepackage{graphicx}

\begin{document}


\title{The Anomalous Skin Effect in Single Crystal Relaxor Ferroelectric
PZN-$x$PT and PMN-$x$PT}


\author{Guangyong Xu$^{1}$, P. M. Gehring$^{2}$, C. Stock$^{3}$, and K. H. Conlon$^{4}$}
\affiliation{
   $^{1}$Condensed Matter Physics and Materials Science Department, Brookhaven National Laboratory, Upton, New York 11973-5000 \\
   $^{2}$NIST Center for Neutron Research, National Institute of Standards and Technology,
   Gaithersburg, Maryland 20899-8562 \\
   $^{3}$Department of Physics and Astronomy, Johns Hopkins University, Baltimore, MD, 21218 \\
   $^{4}$Atomic Energy Canada Limited, Chalk River Nuclear Laboratories, Chalk River Ontario, K0J 1J0 Canada}

\date{\today}

\begin{abstract}

X-ray and neutron scattering studies of the lead-based family of
perovskite relaxors PZN-$x$PT and PMN-$x$PT have documented a highly
unusual situation in which the near-surface region of a single
crystal can exhibit a structure that is different from that of the
bulk when cooled to low temperatures.  The near-surface region, or
"skin" can also display critical behavior that is absent in the
crystal interior, as well as a significantly different lattice
spacing.  By varying the incident photon energy, and thus the
effective penetration depth, x-ray measurements indicate a skin
thickness of order 10~$\mu$m to 50~$\mu$m for PZN-$x$PT samples with
$0 \le x \le 8$\%.  Neutron residual stress measurements on a large
PMN single crystal reveal a uniform lattice spacing within the bulk,
but an increased strain near the surface.  The presence of this skin
effect has led to incorrect phase diagrams for both the PZN-$x$PT
and PMN-$x$PT systems and erroneous characterizations of the nature
of the relaxor state.

\end{abstract}


\maketitle


\newpage

\section{Introduction}

The exceptional piezoelectric properties of the lead-oxide class of
relaxor ferroelectrics has fueled an intense amount of scientific
research on these materials over the past several
years.~\cite{Park,Service} Numerous studies have been performed in
an attempt to understand the fundamental mechanism(s) responsible
for the ultrahigh piezoelectric response of these
compounds.~\cite{Ye_review} During the course of these studies,
additional anomalous behavior has been documented including the
truly remarkable situation in which the structure of the bulk of a
single crystal relaxor specimen can differ from that of the
near-surface region, or "skin," which spans tens of microns.  The
x-ray scattering studies of Xu {\it et al.} have revealed a striking
discrepancy between the low-temperature crystal structure reported
by earlier x-ray scattering studies on the parent compound
Pb(Zn$_{1/3}$Nb$_{2/3}$)O$_{3}$ (PZN) and that observed using
high-energy x-rays.~\cite{Xu_PZN1,Xu_PZN2,Xu_APL}  By varying the
energy of the incident x-rays, and thus the penetration depth,
distinct crystal structures were observed.  Subsequent neutron
scattering studies by Gehring {\it et al.} have found that the same
discrepancy exists in solid solutions of
Pb(Mg$_{1/3}$Nb$_{2/3}$)O$_{3}$ (PMN) mixed with 10\%\ PbTiO$_3$
(PT).~\cite{Gehring_PMN10PT}  More recent neutron residual stress
measurements on a large single crystal of PMN at room temperature
show that the cubic unit cell lattice parameter varies with the
depth normal to the crystal surface, with a corresponding large
surface strain.~\cite{Conlon}

In this paper we review and expand upon the current understanding of
the skin effect in the PZN-$x$PT and PMN-$x$PT relaxor systems.  New
data are reported that demonstrate the high variability in crystal
structure that can be obtained from x-ray measurements made on the
same single crystal specimen, but at different locations over the
crystal surface.  A detailed analysis of high resolution neutron
data obtained on PMN-10\%PT single crystal reveals the presence of
strong internal strain that does not manifest itself as a true
structural transition, but is nonetheless correlated with the
cubic-to-rhombohedral phase transition temperature determined with
low-energy x-rays.  New data is also presented that demonstrates
that the depth dependence of the lattice constant measured in pure
PMN varies significantly with temperature.  While these findings are
of obvious significance from a fundamental research point of view,
they are also important because the skin effect has misled
researchers into drawing incorrect conclusions about the PZN-$x$PT
and PMN-$x$PT phase diagrams, as well as several other erroneous
assumptions about the true relaxor state.

\section{Low Temperature Structure of PZN:  X-Rays}

The phase diagrams of the PZN-$x$PT and PMN-$x$PT systems are
striking in their similarity; both exhibit cubic-to-rhombohedral
phase transitions at low PT content, cubic-to-tetragonal phase
transitions at high PT content, and a steeply-sloped morphotropic
phase boundary (MPB) separating monoclinic~\cite{Noheda_PRL} and
tetragonal phases over a narrow range of intermediate PT
concentrations.  Figure~1 shows the phase diagram for PMN-$x$PT.  In
addition, all compounds below the MPB exhibit exceptionally large
piezoelectric coefficients as well as other properties
characteristic of relaxors, namely a strongly frequency dependent
dielectric susceptibility and a very large room temperature
dielectric constant.  Given these numerous similarities, the $x=0$
case of pure PMN stands out as an oddity because several studies
have shown that the crystal structure retains an average cubic
structure down to 5~K,~\cite{Bonneau, deMathan} whereas the x-ray
measurements by Lebon {\it et al.} show that pure PZN transforms to
a rhombohedral phase at a transition temperature $T_c =
410$~K.~\cite{Lebon}  The unique status of PMN was made all the more
curious after an intriguing study by Ohwada {\it et al.} on a
zero-field cooled single crystal of PZN-8\%PT did not find the
expected rhombohedral phase below $T_c$.~\cite{Ohwada}  These
results motivated the decision by Xu {\it et al.} to re-examine the
zero-field structure of pure PZN, especially because previous
studies of the structure of PZN-$x$PT had been done on poled single
crystals.~\cite{Xu_PZN1}  Another motivating factor were the
measurements by Noheda {\it et al.} that showed that x-ray
diffraction results on PZN-$x$PT single crystals depended strongly
on surface structures.~\cite{Noheda_Ferro, Noheda_PRL}

\begin{figure}
\includegraphics[width=6cm]{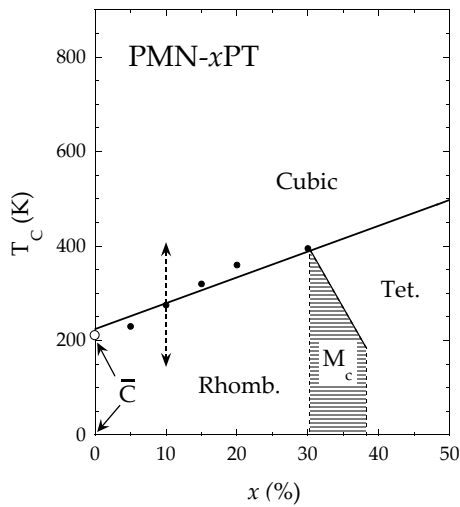}
\caption{ \label{Fig1} Zero-field phase diagram of PMN-$x$PT.}
\end{figure}

Single crystals of PZN were grown via spontaneous nucleation from a
high temperature solution using PbO as flux.~\cite{Zhang}  Two
triangular-shaped plates 1~mm thick and with edges roughly 3~mm in
length were cut with large surfaces parallel to the cubic [111] axis
from the same as-grown single crystal.  The faces of both crystals
were polished using a series of diamond pastes down to 3~$\mu$m, and
gold was sputtered onto the large faces of one crystal, which was
then poled at 200~C and cooled to room temperature in a field of
20~kV/cm.  The unsputtered sample was neither poled nor thermally
treated, thereby leaving a potentially larger mechanical strain
within the crystal. (Our measurements have shown that the structures
of the skin/bulk do not vary after thermal cycling up to 800 K.) The
x-ray measurements on PZN reported here were all performed at the
Brookhaven National Laboratory National Synchrotron Light Source
(NSLS), primarily on beamline X-17B1, which can provide x-rays with
an incident energy of 67~keV and an energy resolution $\Delta E/E =
10^{-4}$.  The spatial extent of the focused x-ray beam on the
sample was $0.5 \times 0.5$~mm$^2$. Additional measurements were
made on beamline X-22A, which provided access to lower incident
x-ray energies of 32 and 10.7~keV.

\begin{figure}
\includegraphics[width=6cm]{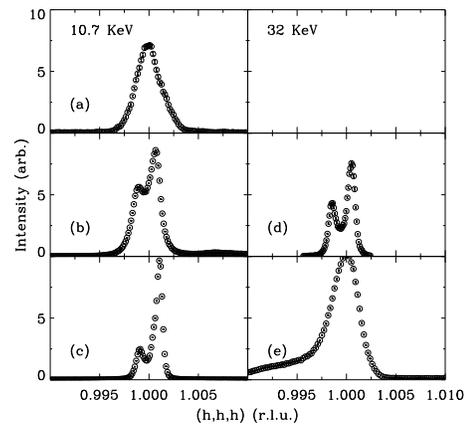}
\caption{ \label{Fig2} Radial scans of the x-ray scattering
intensity measured across the PZN (111) Bragg peak.  Panels on the
left (right) correspond to measurements made with 10.7~keV (32~keV)
incident energy x-rays.  Different panels correspond to different
locations on the crystal surface.}
\end{figure}

Figure~2 shows measurements on the unpoled single crystal of PZN at
room temperature (which is well below the reported value of $T_c =
410$~K) using two different incident energies of 10.7 and 32 keV,
corresponding to penetration depths of 13~$\mu$m and 59~$\mu$m,
respectively.  The different panels in each figure column correspond
to measurements made on different locations of the crystal surface.
The data correspond to radial ($\theta - 2\theta$) scans of the
scattering intensity measured across the (111) Bragg peak, which
must split into two peaks in a rhombohedral phase.  It is
immediately apparent that the different measurements do not confirm
a single crystal structure.  The top-left panel shows a cubic-like
single peak at one location (panel (a)), whereas a two-peaked
structure, suggestive of rhombohedral symmetry, is observed at two
others (panels (b) and (c)).  The same discrepancy holds at the
higher incident x-ray energy of 32~keV (panels (d) and (e)) even
though these data correspond to a deeper penetration depth.

\begin{figure}
\includegraphics[width=6cm]{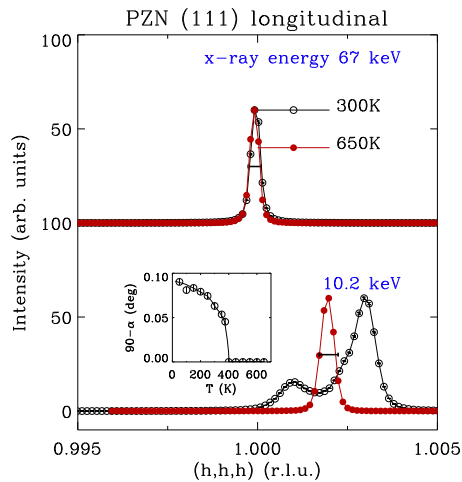}
\caption{ \label{Fig3} Top panel:  radial scans of PZN (111) Bragg
peak measured at 300~K (open circles) and 650~K (closed circles)
using 67~keV incident energy x-rays.  Bottom panel:  radial scans of
PZN (111) Bragg peak measured at 300~K (open circles) and 650~K
(closed circles) using 10.2~keV incident energy x-rays.  Inset shows
temperature dependence of the rhombohedral angle
$\alpha$.~\cite{Xu_PZN2}}
\end{figure}

In order to probe the structure to still greater depths within the
crystal, x-ray measurements were made of the same (111) reflection
at an incident energy of 67~keV, which corresponds to a penetration
depth of 412~$\mu$m.  The upper panel of Fig.~3 shows the resulting
(111) Bragg peak profile, using the same radial scans shown in
Fig.~2, at temperatures above (650~K) and below (300~K) $T_c =
410$~K.~\cite{Xu_PZN2}  Remarkably, a single, resolution-limited
peak is observed at both temperatures thereby indicating the absence
of the rhombohedral phase previously reported by Lebon {\it et al}.
At all different locations measured on the crystal at this energy,
only a single resolution-limited (111) Bragg peak was observed.  An
effort to reproduce the lower energy (8.9~keV from Cu K$_{\beta}$
radiation) results of Lebon {\it et al.} were made by repeating the
same measurements at 10.7~keV.  These data are shown in the lower
panel of Fig.~3 where it is seen that a cubic structure (single
peak) is observed at 650~K as was the case at 67~keV.  At 300~K,
however, a clear split (111) Bragg peak profile is seen, consistent
with a rhombohedral distortion.  The temperature dependence of the
rhombohedral angle $\alpha$ is plotted in the inset to Fig.~3, and
gives a transition temperature of 410~K, also consistent with
previous reports.

\begin{figure}
\includegraphics[width=6cm]{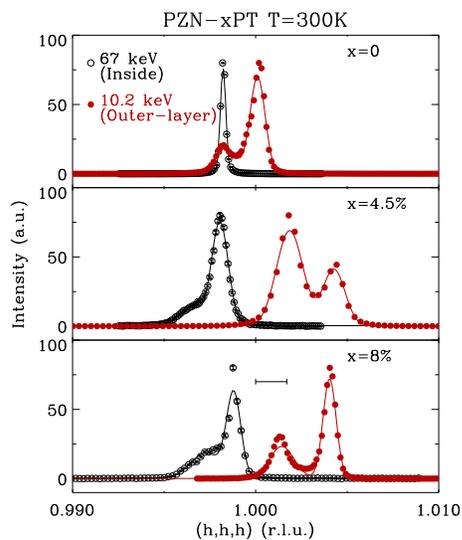}
\caption{ \label{Fig4} Radial scans through the (111) Bragg peak at
300~K for PMN (top panel), PMN-4.5\%PT (middle panel), and PMN-8\%PT
(bottom panel).  Open (closed) circles correspond to an incident
energy of 67~keV (10.2~keV).~\cite{Xu_APL}}
\end{figure}

This combination of x-ray measurements, which probed varying depths
of the PZN near-surface region, provides conclusive evidence for the
existence of a skin effect whereby the bulk and skin of a single
crystal exhibit different structures.  Additional evidence of this
effect was reported in a recent neutron scattering study on PZN by
Stock {\it et al.}, who found that the Bragg peaks broaden
significantly below the x-ray determined value of $T_c = 410$~K
along both the radial and transverse directions, but saw no sign of
a rhombohedral distortion.~\cite{Stock_PZN} Assuming that the phase
diagrams of PZN-$x$PT and PMN-$x$PT should be "universal," the
high-energy x-ray data on PZN reconcile the puzzling discrepancy at
$x=0$ in that they demonstrate that both PMN and PZN retain an
average cubic structure down to low temperatures. This leads
naturally to the question of how the skin is affected by the
addition of PbTiO$_3$ (PT).  To answer this, three single crystals
of PZN-$x$PT with PT concentrations $x=0$, 4.5\%, and 8\%\ were
studied at room temperature by Xu {\it et al.} in the same manner as
described above, i.\ e.\ using incident x-ray energies of 10.2~keV
and 67~keV. The data are shown in Fig.~4.~\cite{Xu_APL} The top
panel summarizes the situation just described for PZN ($x=0$), where
skin and bulk exhibit different crystal structures at 300~K. At a PT
concentration of 4.5\%, however, both the skin region (10.2~keV
data) and the crystal bulk (67~keV data) exhibit a rhombohedral
phase. The difference here is that the bulk rhombohedral phase is
characterized by a smaller rhombohedral angle $\alpha$ than is that
of the skin. Similar results were obtained at a PT concentration of
8\%, just below the MPB, where again both skin and bulk exhibit a
rhombohedral distortion, but with a smaller value of $\alpha$
measured in the bulk.  Given the penetration depth of the 10.2~keV
x-rays, the thickness of the skin was estimated at roughly
10~$\mu$m$-$50~$\mu$m for all samples.

\begin{figure}
\includegraphics[width=6cm]{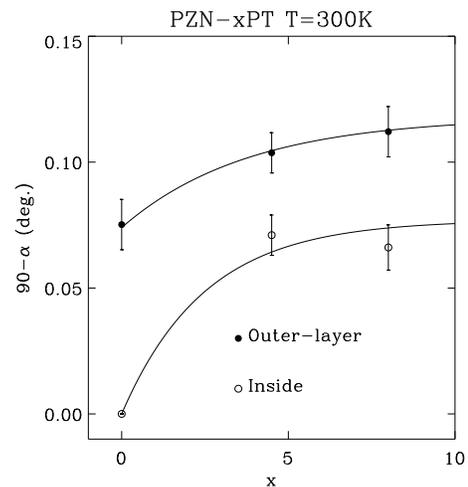}
\caption{ \label{Fig5} Composition dependence of the rhombohedral
angle $\alpha$. Closed circles correspond to the skin, or
outer-layer, measured at 10.2~keV.  Open circles correspond to the
crystal bulk measured at 67~keV.~\cite{Xu_APL}}
\end{figure}

The PT dependence of $\alpha$ for both the outer-layer skin region
and the crystal bulk is given in Fig.~5.~\cite{Xu_APL}  These data
demonstrate that the addition of ferroelectric PT stabilizes a
low-temperature rhombohedral phase that is otherwise absent in pure
PZN, and they provide new insight into the way the structure of the
relaxor PZN evolves as one approaches the MPB where the greatest
piezoelectric response is observed.  The x-ray studies of PZN
reported here have exploited the variability of the penetration
depth by appropriate tuning of the incident x-ray energy.  Neutrons,
by contrast, have penetration depths of order 1~cm, and thus provide
bulk information.  Motivated by the x-ray work on PZN, a program of
neutron scattering measurements on PMN-$x$PT crystals was undertaken
to determine how the rhombohedral phase evolves when adding PT into
pure PMN.

\section{Modified Phase Diagram of PMN-$x$PT:  Neutrons}

The accepted phase diagram for PMN-$x$PT, shown previously in
Fig.~1, indicates that a cubic-to-rhombohedral phase transition
takes place for $x$ less than about 32\%.~\cite{Choi,Noheda_PMN} The
solid data points are values of the transition temperature $T_c$
measured by different authors,~\cite{Ye_PMNxPT} and the solid line
is a linear fit to these data.  The solitary open circle at $x=0$
corresponds to the first-order phase transition that has been
reported to occur in PMN near 213\,K, but only after first cooling
in an electric field $E \ge 1.7$\,kV/cm.~\cite{Ye_review} The
notation $\overline{\rm{C}}$ shown by the arrow is used to denote
the fact that PMN retains an average cubic structure below this
temperature when zero-field cooled (ZFC). A monoclinic (M$_c$) phase
was discovered by Noheda {\it et al.}~\cite{Noheda_PMN} at higher PT
concentrations next to the well-known morphotropic phase boundary
(MPB) separating the M$_c$ phase from the tetragonal phase.

\begin{figure}
\includegraphics[width=6cm]{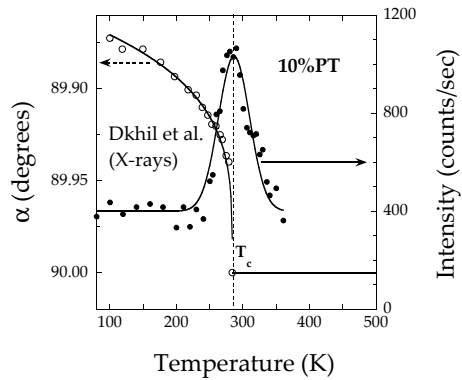}
\caption{ \label{Fig6} Temperature dependence of the rhombohedral
angle $\alpha$ and diffuse scattering measured with x-rays by Dkhil
{\it et al.} on PMN-10\%PT.~\cite{Dkhil} Lines are guides to the
eye.}
\end{figure}

An interesting and detailed comparative study between PMN and
PMN-10\%PT was carried out by Dkhil and co-workers using both x-ray
and neutron scattering techniques on powder and on single crystal
samples.~\cite{Dkhil}  Their results point to the existence of
competing rhombohedral and tetragonal order that never produces a
ferroelectric distortion in PMN, but does in PMN-10\%PT below a
critical temperature $T_c = 285$~K.  In addition to a rhombohedral
phase, critical-like diffuse scattering was observed that peaked at
$T_c$.  The temperature dependence of both this diffuse scattering,
as well as the rhombohedral angle $\alpha$ is given in
Fig.~6.~\cite{Dkhil} Subsequent x-ray scattering experiments by Ye
{\it et al}.\ on a series of PMN-$x$PT samples detected a
low-temperature rhombohedral distortion for PT concentrations as
small as 5\%.~\cite{Ye_PMNxPT} Because all of the published x-ray
and neutron structural data agree in the case of PMN (both indicate
a persistent cubic phase down to 5~K) it was believed that the skin
effect in the PMN-$x$PT system, if it exists at all, must be
significantly smaller than that in PZN-$x$PT. To test this idea, a
single crystal of PMN-10\%PT was obtained for the purpose of
performing a high resolution neutron structural characterization of
the material over a wide range of
temperatures.~\cite{Gehring_PMN10PT}  The dashed vertical arrow in
Fig.~1 locates the structural transition at 10\%\ PT.

The neutron scattering study was performed on the BT9 triple-axis
spectrometer located at the NIST Center for Neutron Research.
Diffuse scattering measurements were performed using fixed incident
and final neutron energies $E_i = E_f = $14.7\,meV ($\lambda =
2.36$\,\AA), which were obtained via Bragg diffraction from the
(002) reflection of highly-oriented pyrolytic graphite (HOPG)
crystals as monochromator and analyzer, and relatively relaxed
horizontal beam collimations of 40$'$-46$'$-S-40$'$-80$'$ (S =
sample).  To maximize the instrumental $q$-resolution, and thus the
sensitivity to subtle structural distortions and changes in lattice
spacing, the spectrometer was operated with a perfect Ge crystal as
analyzer.  More importantly, the $d$-spacings of both the analyzer
and the PMN-10\%PT Bragg reflections were specifically chosen to
match one another as closely as possible. This latter condition
yields a minimum in the instrumental longitudinal
$q$-resolution.~\cite{Emilio, Xu_highq} For the perovskite
PMN-10\%PT sample the $d$-spacings associated with the (111) and
(022) Bragg peak (2.333\,\AA\ and 1.4287\,\AA, respectively) are
well-matched with those for Ge (220) and (004) (2.000\,\AA\ and
1.4143\,\AA).  Our choice of horizontal beam collimations of
10$'$-46$'$-S-20$'$-40$'$ was based on Monte Carlo intensity
simulations to improve the $q$-resolution without an unreasonable
reduction in intensity.~\cite{Vinita} An incident neutron energy of
8.5\,meV was used for measurements of the (111) Bragg peak, which
was provided via a highly oriented pyrolytic graphite monochromator,
and resulted in a sharp $q$-resolution of 0.0018\,rlu FWHM (1\,rlu =
2$\pi/$a = 1.5576\,\AA$^{-1}$).  The measurements of the (022) Bragg
peak yielded an even better $q$-resolution approaching 0.0010\,rlu
FWHM, thanks to a more closely matched $d$-spacing condition between
sample and analyzer.

A 2.65~g single crystal of PMN-10\%PT, with approximate dimensions
of $4 \times 9 \times 11$~mm$^3$, was produced using a top-seeded
solution growth technique.~\cite{Ye_Growth} The crystal mosaic
full-width at half-maximum (FWHM) measured at the (200) Bragg peak
at 500~K is less than 0.06$^{\circ}$ indicating an excellent quality
sample. Access to Bragg reflections of the form $(hhh)$ and $(0ll)$
were needed to study the rhombohedral phase; therefore the crystal
was mounted with the [01$\overline{1}$] axis aligned vertically. The
sample was then mounted inside a high-temperature closed-cycle
$^3$He refrigerator and fixed to the goniometer of the BT9
triple-axis spectrometer.

\begin{figure}
\includegraphics[width=6cm]{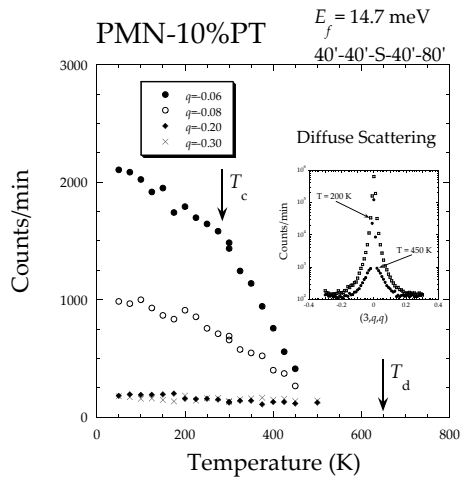}
\caption{ \label{Fig7} Neutron diffuse scattering intensity measured
as a function of temperature at $\vec{Q} = (3,q,q)$ on a PMN-10\%PT
single crystal.  The inset shows the full diffuse scattering profile
measured along the [011] direction (transverse to $\vec{Q}$) at
200\,K and 450\,K.~\cite{Gehring_PMN10PT}}
\end{figure}

The neutron diffuse scattering data on PMN-10\%PT were measured near
the (300) Bragg peak (along the [011] direction) where it is known
that such scattering is very strong.~\cite{Hirota}  The temperature
dependence of the diffuse scattering is shown in Fig.~7 at several
selected values of the reduced wavevector
$q$.~\cite{Gehring_PMN10PT} An immediate discrepancy is apparent
between these and the x-ray data of Dkhil {\it et al}.\ shown in
Fig.~6 inasmuch as the neutron diffuse scattering data do not
exhibit a peak at any temperature down to 50~K. Instead, a slight
suggestion of a change in slope is observed in the vicinity of $T_c
= 285$~K for $q = 0.06$~rlu.  The inset to Fig.~7 shows two full
scans of the (300) peak at 200~K and 450~K showing the strong growth
of the diffuse scattering intensity with decreasing temperature. The
presence of strong diffuse scattering below $T_c$ directly
contradicts the x-ray results, which were obtained using an incident
x-ray energy of 8.9~keV.  This finding strongly suggests that a skin
effect is also present in the PMN-$x$PT system as it provides a
natural explanation with which to reconcile the neutron and x-ray
measurements.

\begin{figure}
\includegraphics[width=6cm]{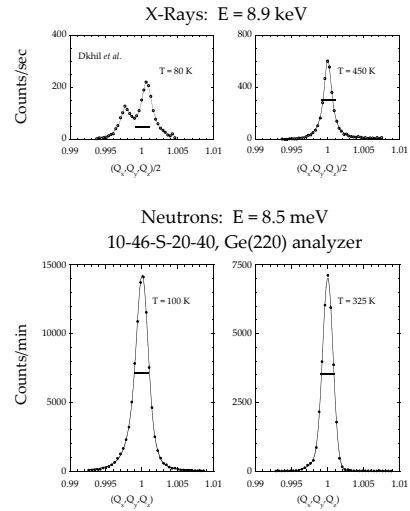}
\caption{ \label{Fig8} Comparison of x-ray measurements by Dkhil
{\it et al.} at (222) (top)~\protect\cite{Dkhil} and our neutron
measurements at (111) (bottom) on PMN-10\%PT. The horizontal scale
of the x-ray data have been reduced by a factor of two to allow
direct comparison with the neutron (111) data.  Solid bars indicate
the BT9 instrumental $q$-resolution FWHM.~\cite{Gehring_PMN10PT}}
\end{figure}

Neutron scattering measurements were made at both the (111) and
(022) Bragg reflections of the PMN-10\%PT sample using the high
$q$-resolution configuration of the BT9 spectrometer described
earlier.  All reflections of the form $(hhh)$ and $(0ll)$ must split
into two in the presence of a rhombohedral distortion because the
unit cell is elongated along one diagonal and compressed along the
others, thus giving rise to two distinct $d$-spacings.  Hence (111)
and $(1\overline{11})$ Bragg peaks will no longer be identically
situated in reciprocal space, and their relative intensities will
give the relative domain populations for each reflection.  Radial
scans of the neutron scattering intensity across the (111) Bragg
peak are presented in Fig.~8 both above and below $T_c=285$~K, and
these are compared to equivalent scans across the (222) Bragg peak
measured with x-rays.~\cite{Gehring_PMN10PT} The horizontal scales
of the x-ray data have been adjusted to allow for a direct scan by
scan comparison.

At high temperatures, both neutron and x-ray data show a single
resolution-limited Bragg peak consistent with a cubic structure. But
below $T_c$ the x-ray measurements reveal a definitive splitting of
the (222) Bragg peak whereas the neutron data do not.  Instead, the
neutron measurements continue to show a single peak.  Moreover, the
neutron peak width is less than the splitting observed in the x-ray
scan, thus providing even stronger confirmation of the absence of
any rhombohedral distortion in the bulk of the PMN-10\%PT single
crystal.  Some broadening is apparent in the neutron case at 100~K.
This could either be the result of internal strain, which would
manifest as a radial broadening, and/or it could be due to the
contribution from the skin.  Spatially resolved neutron diffraction
measurements of the near-surface strain by Conlon {\it et al.} in a
single crystal of pure PMN, where both a dramatic change in the
lattice constant and the Bragg peak intensities is observed over a
length scale of order 100~$\mu$m,~\cite{Conlon} suggest that strain
is the most likely source of this broadening, and will be discussed
in the following section.

\begin{figure}
\includegraphics[width=6cm]{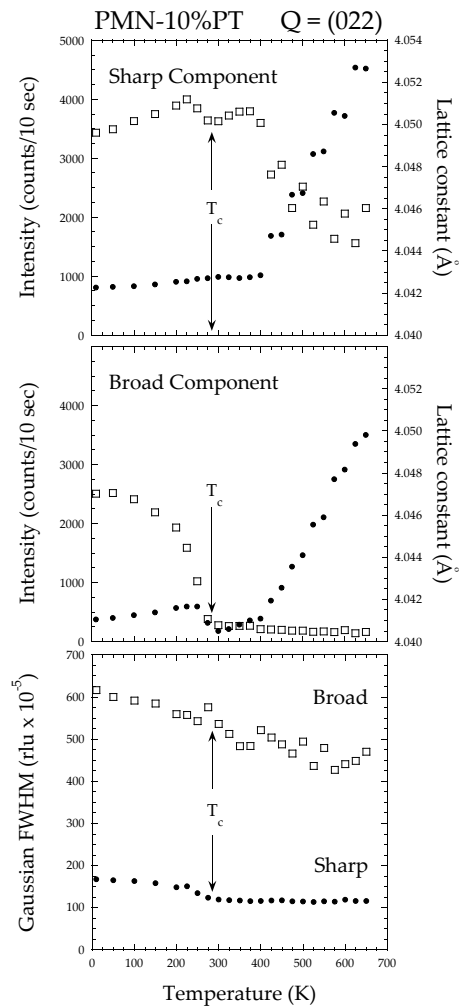}
\caption{ \label{Fig9} Temperature dependence of the two component
fitting analysis for the (022) Bragg peak measured with high
$q$-resolution.  Top and middle panels compare the intensity and }
\end{figure}

The thermal expansion of the PMN-10\%PT sample was measured from
10~K to 600~K by carefully measuring the (022) Bragg peak radial
profile with high $q$-resolution.  It was noted that these data were
significantly better fit using the sum of two Gaussian functions of
$q$, one sharp and the other broad.  The temperature dependences of
the resulting peak intensity and lattice spacing for these two
components are shown in the top and middle panels of Fig.~9.  The
notable feature here is that the sharp component intensity and
associated lattice constant show no significant feature that
correlates with $T_c$.  By contrast, the same parameters associated
with the broad component correlate quite well with the x-ray
determined value of $T_c$.  In particular, the thermal expansion
shown in the middle panel behaves as expected for a normal
ferroelectric.~\cite{Xu_PMNxPT}  We emphasize here that the broad
component is a single peak, and still quite narrow, and thus does
not represent a rhombohedral distortion.  Rather the broad peak is
believed to represent a highly strained region of the crystal bulk.

The lattice constant associated with the sharp component is
intriguing in that it exhibits an almost temperature independent
behavior from 10~K to 400~K, with the $d$-spacing changing by less
than 0.001~\AA\ over this temperature range.  However at 400~K both
the lattice constant and the peak intensity exhibit abrupt changes.
Indeed, the rate of thermal expansion becomes extremely large, being
roughly $10^{-5}$~K$^{-1}$, which is considerably larger than the
value $10^{-6}$~K$^{-1}$, which is typical of other oxides. It is
suggestive that this temperature is very close to those observed in
pure PMN and PMN-20\%PT, where both the maximum broadening of the TA
mode occurs and the minimum value of the TO soft mode is
obtained.~\cite{Wakimoto_PMN1, Wakimoto_PMN2, Koo}  The bottom panel
of Fig.~9 shows the variation of the peak widths (FWHM) for the two
Gaussian components. In this case it is interesting that the sharp
component FWHM appears to correlate more obviously with $T_c$ than
does that of the broad component.

\section{Depth-Dependent Lattice Parameter and Surface Strain in PMN:  Neutrons}

The x-ray and neutron results obtained on PMN-10\%PT, when viewed
together, point strongly to the presence of a skin effect like that
documented by Xu {\it et al.} in PZN.  The concept of a near-surface
region having a structure distinct from that of the bulk would also
provide a natural explanation of the results from x-ray and neutron
powder diffraction studies of the low-temperature structure of both
PMN and PZN, for which a two-phase model, one cubic and the other
rhombohedral, is required to describe the observed powder peak
intensity profiles.~\cite{Iwase, deMathan} Because neutrons interact
weakly with matter (typical penetration depths being of order 1~cm)
any observable scattering signal from the skin would imply a
substantial skin volume.  This could be the result of a "thick" skin
of order many microns in single crystals, or a large effective
surface area, as would be the case in powdered samples.

\begin{figure}
\includegraphics[width=6cm]{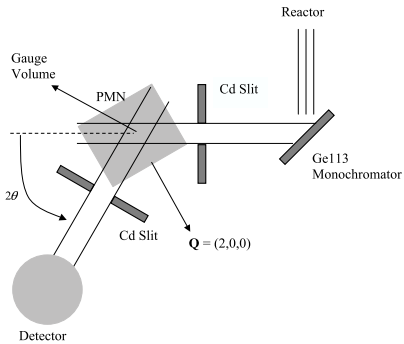}
\caption{ \label{Fig10} Schematic diagram of the neutron residual
stress instrument used for spatially resolved measurements of
strain.  The gauge volume is that region of the crystal bounded by
the intersection of the incident and scattered neutron beams, which
are given by the solid lines  The slits on the incident and
scattered beams were 5~mm tall and 300~$\mu$m wide.~\cite{Conlon}}
\end{figure}

To examine this idea, neutron residual stress measurements were
performed by Conlon {\it et al.} on the L3 double-axis spectrometer,
which is located at the National Research Universal (NRU) reactor at
Chalk River Laboratories, in order to study neutrons diffracted from
a very large 9.3~cc single crystal of PMN as a function of
depth.~\cite{Conlon} The PMN crystal was grown by the modified
Bridgeman technique, the details of which are described
elsewhere.~\cite{Luo}  A schematic diagram of the L3 instrument is
shown in Fig.~10, where a Ge(113) crystal was used to monochromate
the incident beam to an energy of $E_i = 14.5$~meV, and horizontal
beam collimations of 60$'$-30$'$-S-30$'$ were used in conjunction
with cadmium (strong neutron absorbers) slits placed both before and
after the sample. An HOPG filter was placed in the incident beam to
remove higher order harmonics (i.\ e.\ neutrons with wavelengths
$\lambda/2$, $\lambda/3$, etc.).  The crystal was cut in order to
provide a large and well-defined {100} surface, and neutrons
scattered from the (200) Bragg peak from this surface at room
temperature were recorded in a reflection geometry as indicated in
the schematic.

The residual stress technique allows a measurement of lattice strain
as a function of depth normal to the crystal surface through the use
of narrow absorbing slits, or masks, that provide a high degree of
angular collimation of both the incident and scattered beams.  The
result is that only neutrons scattered from a small "gauge" volume,
represented by the diamond-shaped region in Fig.~10, are recorded by
the detector.  By translating the large single crystal along the
scattering vector $\vec{Q}$, in this case the (200) Bragg peak, and
then measuring radial scans of the Bragg peak intensity at each
value of translation, the lattice constant of the sample can be
probed as a function of depth.

\begin{figure}
\includegraphics[width=6cm]{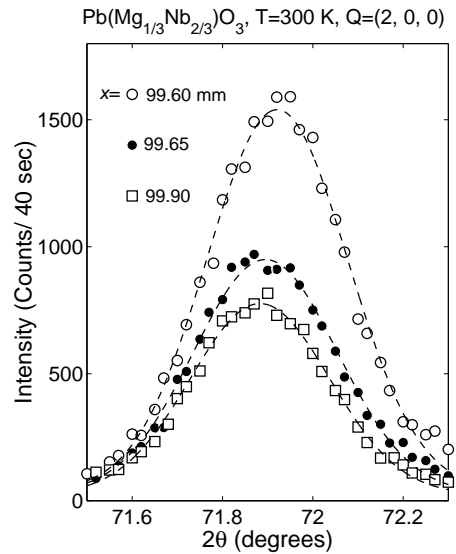}
\caption{ \label{Fig11} Radial ($\theta-2\theta$ scans of the (200)
Bragg peak scattering intensity are shown for three values of
translation of the PMN crystal along the [100] axis
 Dashed lines are fits to a Gaussian lineshape.~\cite{Conlon}}
\end{figure}

Data from $theta-2\theta$ scans are shown in Fig.~11 for three
different translational settings.~\cite{Conlon}  Two features are
readily apparent from this figure:  the first is the shift of the
center of the (200) Bragg peak in $2\theta$, thus indicating a
measurable change of the lattice parameter over a depth of 0.3~mm
(300~$\mu$m), while the second is the dramatic change in the peak
intensity.  This latter effect is due to the extinction effect,
which has been observed previously in analogous neutron scattering
measurements of the second length scale phenomenon in Tb and
SrTiO$_3$ for which a residual stress configuration was
used.~\cite{Gehring_Tb1, Hirota_Tb1, Gehring_Tb2, Rutt}  As a cross
check of these data, a perfect germanium single crystal was studied
in exactly the same way using the Ge (220) reflection.  The
$d$-spacing of the Ge (220) Bragg peak is almost identical to that
of the PMN (200) Bragg peak. This allows the instrument scattering
angles to remain nearly the same, and thus provides a direct test of
the observed change in lattice parameter with depth (translation).
Unlike the case of PMN, no significant strain or change in lattice
spacing was observed in the near surface region of the perfect
germanium single crystal over a depth of more than 1000~$\mu$m.  An
additional check was made by measuring rocking curves ($\theta$
scans) to determine the crystal mosaic as a function of depth.  The
mosaics obtained near the surface and deep within the bulk were both
measured to be $0.185 \pm 0.007^{\circ}$, thus demonstrating that
the change in lattice constant is not an artifact of a variation in
mosaic spread or the presence of another crystal domain near the
surface.

\begin{figure}
\includegraphics[width=6cm]{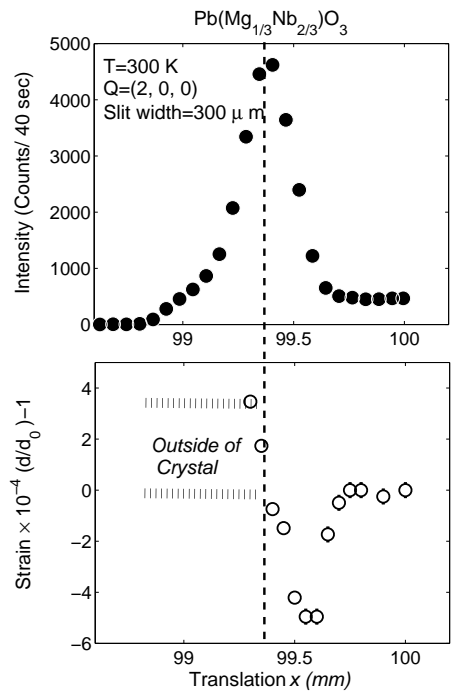}
\caption{ \label{Fig12} Top panel:  (200) Bragg peak intensity as a
function of translation (depth).  Bottom panel:  normalized lattice
spacing (strain) $(d/d_0)-1$ as a function of translation, where
$d_0$ is the bulk value of the lattice spacing $d$.  The location of
the crystal surface is indicated by the vertical dashed
line.~\cite{Conlon}}
\end{figure}

Figure~12 shows the depth dependence of the (200) Bragg peak
intensity (top panel) and the crystal lattice spacing (bottom panel)
over a range exceeding 500~$\mu$m.~\cite{Conlon}  The large
variation in the peak intensity is due to the extinction effect
mentioned above, and cannot be ascribed solely to the linear
attenuation of the incident and scattered neutron beams as they
traverse through the PMN crystal.  The variation of Bragg intensity
is important, however, as it can be used to locate the surface of
the sample on the translation axis.  The crystal surface should
correspond to the translation that gives the maximum intensity as
this will be when the gauge volume lies completely below the crystal
surface.  At this point, the scattering volume is a maximum, while
the effects from extinction are minimal. The bottom panel shows a
strong variation in the lattice spacing spanning several hundred
microns (see Fig.~11), and thus confirms the existence of a skin
effect in pure PMN.  It thus appears to be the case that the skin is
not "thin" in PMN; rather the crystal structures of the bulk and
skin are both cubic, and there is a substantial near surface strain.

\begin{figure}
\includegraphics[width=6cm]{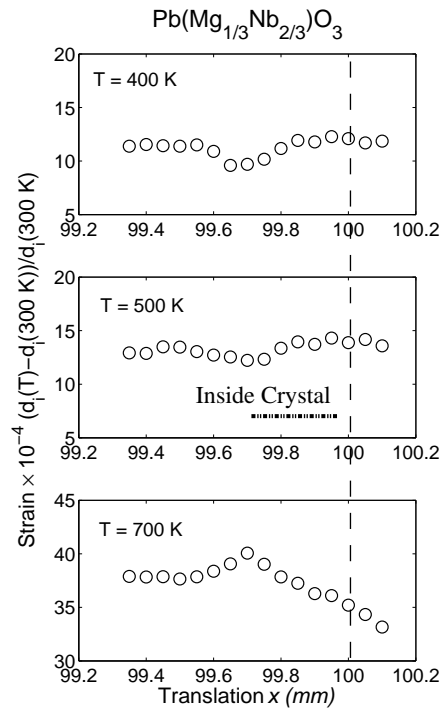}
\caption{ \label{Fig13} Normalized lattice spacing, or strain,
$(d(T)/d_i(300$~K$)-1$ as a function of translation, measured at
400~K, 500~K, and 700~K.  Here $d_i(300$~K) is the bulk value of the
lattice spacing $d$ at room temperature. The location of the crystal
surface is indicated by the vertical dashed line.}
\end{figure}

All of the residual stress measurements presented so far were
performed at room temperature.  Figure~13 shows data obtained in a
manner identical to that in the bottom panel of Fig.~12, but at
three higher temperatures, namely 400~K, 500~K, and 700~K.  The
striking aspect of these data is the reversal of the strain minimum
apparent at 300~K in Fig.~12, which slowly evolves into a maximum at
700~K in the bottom panel of Fig.~13.  While the number of
temperatures reported here are insufficient to draw any definitive
conclusions about this effect, it is very interesting to note that
the Burns temperature $T_d$, below which local regions of randomly
oriented polar order on a nanometer scale first appear,~\cite{Burns}
lies between 600~K and 700~K.  It is tempting to speculate that
these polar nanoregions might have some relationship to the skin,
and that they may also play an important role in the invar-like
thermal expansion observed from 10~K to 400~K.

\section{Discussion}

The combination of x-ray and neutron measurements presented here
provide conclusive proof of a significant surface layer in single
crystal specimens of both PZN-$x$PT and PMN-$x$PT systems.  The
presence of this skin has led to incorrect phase assignments in the
phase diagrams of both systems inasmuch as no rhombohedral phase is
present at sufficiently low PT concentrations in either case, the
low-temperature phase remaining cubic.  A crossover to a
low-temperature rhombohedral phase does take place in PZN-$x$PT
between $x = 0$ and $x = 4.5$\%\, whereas a similar crossover occurs
in the PMN-$x$PT system between $x = 20$\%\ and $x =
27$\%\.~\cite{Xu_PMNxPT}  That the skin effect persists over such a
large temperature range, and that the skin thickness spans many tens
of microns, are extremely unusual. In the case of the second length
scale phenomena observed in Tb and SrTiO$_3$, the anomalous extra
component in the critical scattering associated with a phase
transition was also observed over a large depth of order 100~$\mu$m,
but it persisted only over a very limited temperature
range.~\cite{Gehring_Tb1,Rutt} In some of these relaxor compounds,
the skin and crystal bulk could exhibit different crystal structures
(PZN and PMN-10\%PT), while in others they possess the same
structure (cubic or rhombohedral), but with either different lattice
spacings (PMN) or different sized rhombohedral distortions
(PZN-4.5\%PT and PZN-8\%PT).  Some logical questions for future
study are whether or not the skin varies with electric field? If so
does it get bigger or smaller?  Such studies are in progress,
including one to examine the field-induced transition in PMN with
high-$q$ resolution, as well as the effects on the strong diffuse
scattering.~\cite{Stock_PMN} It is evident that future theoretical
models of these relaxor systems will need to account for the
observed anomalous skin effect.

\begin{acknowledgments}

The authors wish to acknowledge first and foremost the friendship
and support provided by Dr. Gen Shirane of Brookhaven National
Laboratory, who passed away suddenly in January of 2005.  His
guidance, genius, and boundless enthusiasm formed the foundation for
all of the relaxor studies presented here.  Gen's unceasing
scientific curiosity and his tireless drive to discover new physical
phenomenon were a continuous source of inspiration to all of us.  He
shall be sorely missed.  We dedicate this paper to his memory, and
we express our deep gratitude for all that he has taught us.  The
authors also wish to acknowledge stimulating discussions with Dwight
Viehland and JieFang Li.  Finally, financial support from the U.S.
Department of Energy under Contract DE-AC02-98CH10886 is also
gratefully acknowledged.

\end{acknowledgments}



\begin{thebibliography}{99}

\bibitem{Park} S.-E. Park and T. R. Shrout, J. Appl. Phys. {\bf 82}, 1804 (1997).

\bibitem{Service} R. F. Service, Science {\bf 275}, 1878 (1997).

\bibitem{Ye_review} Z.-G. Ye, {\it Key Engineering Materials Vols. 155-156}, 81 (1998).

\bibitem{Xu_PZN1} G. Xu, Z. Zhong, Y. Bing, Z.-G. Ye, C. Stock, and G. Shirane, Phys. Rev. B {\bf 67}, 104102 (2003).

\bibitem{Xu_PZN2} G. Xu, Z. Zhong, Y. Bing, Z.-G. Ye, C. Stock, and G. Shirane, Phys. Rev. B{\bf 70}, 064107 (2004).

\bibitem{Xu_APL} G. Xu, H. Hiraka, G. Shirane, and K. Ohwada, Appl. Phys. Lett. {\bf 84}, 3975 (2004).

\bibitem{Gehring_PMN10PT} P. M. Gehring, W. Chen, Z.-G. Ye, and G. Shirane, J. Phys.: Condens. Matter {\bf 16}, 7113 (2004).

\bibitem{Conlon} K. H. Conlon, H. Luo, D. Viehland, J. F. Li, T. Whan, J. H. Fox, C. Stock, and G. Shirane, Phys. Rev. B {\bf 70}, 172204 (2004).

\bibitem{Noheda_PRL} B. Noheda, D. E. Cox, G. Shirane, S.-E. Park, L. E. Cross, and Z. Zhong, Phys. Rev. Lett. {\bf 86}, 3891 (2001).

\bibitem{Bonneau} P. Bonneau, P. Garnier, E. Husson, and A. Morell, Mater. Res. Bull. {\bf 24}, 201 (1989); P. Bonneau, P. Garnier, G. Calvarin, E. Husson, J. R. Gavarri, A. W. Hewat, and A. Morell, J. Solid State Chem. {\bf 91}, 350 (1991).

\bibitem{deMathan} N. de Mathan, E. Husson, G. Calvarin,  J. R. Gavarri, A. W. Hewat, and A. Morell, J. Phys.: Condens. Matter {\bf 3}, 8159 (1991).

\bibitem{Lebon} A. Lebon, H. Dammak, G. Calvarin, and I. Ould Ahmedou, J. Phys.:  Condens. Matter {\bf 14}, 7035 (2002).

\bibitem{Ohwada} K. Ohwada, K. Hirota, P. W. Rehrig, Y. Fuji, and G. Shirane, Phys. Rev. B {\bf 67}, 094111 (2003).

\bibitem{Noheda_Ferro} B. Noheda, D. E. Cox, and G. Shirane, Ferroelectrics {\bf 267}, 147 (2002).

\bibitem{Zhang} L. Zhang, M. Dong, Z.-G. Ye, Mater. Sci. Eng., B {\bf 78}, 96 (2000).

\bibitem{Stock_PZN} C. Stock, R. J. Birgeneau, S. Wakimoto, J. S. Gardner, W. Chen, Z.-G. Ye, and G. Shirane, Phys. Rev. B {\bf 69}, 094104 (2004).

\bibitem{Choi} S. W. Choi, T. R. Shrout, S. J. Jang, A. S. Bhalla, Ferroelectrics {\bf 100}, 29 (1989).

\bibitem{Noheda_PMN} B. Noheda, D. E. Cox, G. Shirane, J. Gao, and Z.-G. Ye, Phys. Rev. B {\bf 66}, 054104 (2002).

\bibitem{Ye_PMNxPT} Z.-G. Ye, Y.Bing, J. Gao, A. A. Bokov, P. Stephens, B. Noheda, and G. Shirane, Phys. Rev. B {\bf 67}, 104104 (2003).

\bibitem{Dkhil} B. Dkhil, J. M. Kiat, G. Calvarin, G. Baldinozzi, S. B. Vakhrushev, and E. Suard, Phys. Rev. B {\bf 65}, 024104 (2001).

\bibitem{Emilio} J. E. Lorenzo {\it et al}., Phys. Rev. B {\bf 50}, 1278 (1994).

\bibitem{Xu_highq} G. Xu, P. M. Gehring, V. J. Ghosh, and G. Shirane, Acta Cryst. A {\bf 60}, 598 (2004).

\bibitem{Vinita} Our choice of beam collimations were based on Monte Carlo intensity simulations performed by Vinita Ghosh.

\bibitem{Ye_Growth} Z.-G. Ye, P. Tissot, and H. Schmid, Mater. Res. Bull. {\bf 25}, 739 (1990).

\bibitem{Hirota} K. Hirota, Z.-G. Ye, S. Wakimoto, P. M. Gehring, and G. Shirane, Phys. Rev. B \textbf{65}, 104105 (2002).

\bibitem {Wakimoto_PMN1} S. Wakimoto, C. Stock, R. J. Birgeneau, Z.-G. Ye, W. Chen, W. J. L. Buyers, P. M. Gehring, and G. Shirane, Phys. Rev. B {\bf 65}, 172105 (2002).

\bibitem{Wakimoto_PMN2} S. Wakimoto, C. Stock, Z.-G. Ye, W. Chen, P. M. Gehring, and G. Shirane, Phys. Rev. B {\bf 66}, 224102 (2002).

\bibitem{Koo} T. Y. Koo, P. M. Gehring, G. Shirane, V. Kiryukhin, S.-G. Lee, S.-W. Cheong, Phys. Rev. B {\bf 65}, 144113 (2002)

\bibitem{Iwase} T. Iwase, H. Tazawa, K. Fujishiro, Y. Uesu, and Y. Yamada, J. Phys. Chem. Solids {\bf 60}, 1419 (1999).

\bibitem{Luo} H. Luo, G. Xu, H. Xu, P. Wang, and Z. Yin, Jpn. J. Appl. Phys, Part 1 {\bf 39}, 5581 (2000).

\bibitem{Gehring_Tb1} P. M. Gehring, K. Hirota, C. F. Majkrzak, and G. Shirane, Phys. Rev. Lett. {\bf 71}, 1087 (1993).

\bibitem{Hirota_Tb1} K. Hirota, G. Shirane, P. M. Gehring, and C. F. Majkrzak, Phys. Rev. B {\bf 49}, 11967 (1994).

\bibitem{Gehring_Tb2} P. M. Gehring, K. Hirota, C. F. Majkrzak, and G. Shirane, Phys. Rev. B {\bf 51}, 3234 (1995).

\bibitem{Rutt} U. R\"{u}tt, A. Diederichs, J. R. Schneider, and G. Shirane, Europhys. Lett. {\bf 39}, 395 (1997).

\bibitem{Burns} G. Burns and F.~H.~Dacol, Solid State Commun. \textbf{48}, 853 (1983).

\bibitem{Xu_PMNxPT} G. Y. Xu, D. Viehland, J. F. Li, P. M. Gehring, and G. Shirane, Phys. Rev. B {\bf 68}, 212410 (2003).

\bibitem{Stock_PMN} C. Stock, G. Y. Xu, P. M. Gehring, H. Luo, and G. Shirane, in preparation.

\end{thebibliography}

\newpage

\end{document}